\documentclass[12pt]{article}
\newcommand{\xte}{{\it RXTE\/}}
\newcommand{\bep}{{\it BeppoSAX\/}}
\newcommand{\heao}{{\it HEAO-1\/}}
\newcommand{\ginga}{{\it Ginga\/}}
\usepackage{aasms4}
\usepackage{psfig}
\usepackage{rotating}
\begin{document}

\title{Stability of the Cyclotron Resonance Scattering Feature in Her~X-1 with RXTE}
\author{D. E. Gruber, W. A. Heindl, R. E. Rothschild, W. Coburn}
\affil{Center for Astrophysics \& Space Sciences, University of California,
San Diego 92093}
\authoremail{dgruber@mamacass.ucsd.edu}
\author{R. Staubert, I. Kreykenbohm, and J. Wilms}
\affil{Institut f\"{u}r Astronomie und Astrophysik, Abt. Astronomie, 
Universit\"{a}t T\"{u}bingen, D-72076 T\"{u}bingen, Germany}
\authoremail{staubert@astro.uni-tuebingen.de}

\begin{abstract}
Five observations of the hard X-ray spectrum of Her X-1 from \xte\ show 
  that the $\sim \! 41$ keV energy of the cyclotron scattering line is 
  constant within statistics of a few percent per observation.  The overall 
  spectral shape, on the other hand, varies somewhat, with an RMS of 2\%.  
  If the 41 keV feature truly originates as cyclotron resonance scattering 
  in an unchanging 3 $\times$ 10$^{12}$ Gauss dipole field not 
  far above the neutron star surface, these observations constrain the 
  average height of scattering to within a range of 180 meters. This is 
  consistent with models which put the radiating structure within meters 
  of the surface of the neutron star.  In other pulsars observed line 
  centroid changes have been correlated with luminosity changes, and if 
  interpreted as variations of the height at which scattering takes 
  place, many hundreds of meters are required.  These \xte\ data, which
  sample nearly a factor of two in unabsorbed luminosity, are in 
  conflict with a particular model for such an extended radiating 
  structure.  Comparison with other observations over many years
  indicates strongly that the centroid energy of this absorption line has 
  increased some time between 1991 and 1993 by 23\%, from 34 keV to 41 keV.  
  Moreover, the cutoff energy of the spectral continuum increased at the 
  same time from 16 keV to 20 keV, which is, within the statistical error
  of 5\%, in direct proportion to the centroid.  This may be a sign that
  both these characteristics of the spectrum are controlled in the same
  way by the magnetic field strength in the region of scattering.
 
\end{abstract}


\section{Introduction}

The accreting X-ray pulsar, Hercules X-1, emits a spectrum with a
localized depression at roughly 40 keV which is generally interpreted
as an absorption line resulting from scattering of photons on electrons
whose motions are constrained in one spatial dimension to transitions
between Landau levels in a teraGauss magnetic field at the neutron
star's polar cap. Commonly referred to as a cyclotron line, this
feature was discovered in 1977 (Tr\"{u}mper et al. 1978), although
first identified phenomenologically as an emission feature at somewhat
higher energy.  This feature in Her X-1 has since been observed repeatedly 
(e.g. Gruber et al. 1980, Voges 1984, Tueller et al.  1984, Soong et al.
1988, Mihara 1995, Kunz 1995, Dal Fiume et al. 1998).  Similar features
have also been identified in the hard X-ray spectra of many of the
accreting X-ray pulsars (e.g. Mihara 1995, Heindl et al. 1999, Dal
Fiume et al. 2000).  Cyclotron lines afford a diagnostic of plasmas
under extreme conditions and can provide a key to the structure of the
magnetosphere and the mass flow.

Mihara et al. (1998) have reported a study of repeated observations
with \ginga\ of cyclotron lines in a number of X-ray pulsars.  In a
few instances, although not with Her X-1, they found a correlation with
X-ray luminosity which was consistent with the cyclotron scattering
taking place from hundreds to many hundreds of meters above the neutron
star surface, at the peak of an ``accretion mound" structure
proposed by Burnard, Arons \& Klein (BAK, 1991).  BAK predict a strong
dependence of this height on accretion flow.  Thus in this model
variability of the cyclotron line energy can also be an important
diagnostic of the scattering region and accretion process.

Studies of line formation (M\'{e}sz\'{a}ros \& Nagel 1985, Isenberg,
Lamb \& Wang 1998, Araya-Gochez \& Harding 2000) have indicated the 
possibility of line profiles
which deviate from the simple Gaussian and Lorentzian profiles which
have been used for analysis to date.  Such deviations are evident as
particularly broad lines (Heindl 1999, Orlandini 1999). However, the
line of Her X-1 is narrower, and measurement of structure beyond the
first and second moments (centroid and width) requires detector
resolution higher than the $\delta E/E$ of a few obtainable with
inorganic scintillators, as well as very large collecting area and
observing time.

Using data from five observations over two years with the large-area
instruments aboard Rossi X-Ray Timing Explorer (\xte ) we have
performed detailed spectroscopy of the spectral continuum, the
cyclotron line and the fluorescent iron line of Her X-1.  We find only
minor variability between these observations.  
Exercising
careful data selection, we have compared the results with those from
earlier measurements between 1977 and 1994.  For this purpose the
\heao\ data were freshly analyzed with modern software and spectral
models. To a level of perhaps 10\% the pre-1991 spectral shapes are
consistent with each other, as are the spectra from 1994 onwards.
Between 1991 and 1994 there strongly appears to have been a change of
the cyclotron line centroid and the cutoff energy of the continuum.
This is in marked contrast to the power law index and exponential fold
energy, for which the average values remained constant to within a few
percent.  In \S 2 we describe
instrumentation and observations, in \S 3 analysis and results, and in
\S 4 we briefly discuss possible interpretations.  Preliminary results
from a portion of this \xte\ data set have been reported earlier
(Gruber et al. 1997, 1998, 1999).

\section{Instruments and Observations}

The \xte, launched late in 1995 into low earth orbit,  performs
pointed observations with the Proportional Counter Array (PCA, Jahoda
et al.  1996) and the High Energy X-Ray Timing Experiment (HEXTE,
Rothschild et al.  1998). The PCA consists of five Xenon proportional
counters (called PCU's) sensitive in the energy range 2--60 keV with
1400 cm$^2$ each.  The HEXTE consists of eight NaI scintillator
detectors of area 200 cm$^2$ each and energy range 15--250 keV.  Both
instruments are non-imaging and have coaligned fields of view collimated
to 1$^\circ $ FWHM.  Time resolution was microseconds.  Energy resolution
was 16\% at 6 keV for PCA, and 15\% at 60 keV for HEXTE.

The five \xte\ observations (Table 1) each consisted of several
satellite orbits, each orbit typically containing 3300 seconds
on-source. All five PCU's  were used except for the observations of
1996 August, for which only four detectors were operated.  After the
first observation one of the eight HEXTE detectors was lost for
spectroscopic purposes.  The observations were collected over two years
from 1996 and 1997.  Each was selected to be near the peak of the Her
X-1 ``main-on'' state and to avoid periods of eclipse and other forms
of obscuration, such as pre-eclipse dips.  Exposures were varied, but
averaged 15000 s per observation.

The study of long-term changes relies particularly on archival
\heao\ A-4 (Matteson 1978) data, which we re-analyzed with modern
software and spectral models.  The two Low-Energy Detectors of this
experiment each had area 103 cm$^2$ and spectral resolution 25\% at 
60 keV.  The A-4 instrument was quite accurately calibrated.  The
calibration is described in part in Jung (1986) and Gruber et al.
(1989), and additional relevant details are given in the Appendix.

\section{Analysis and Results}

The spectral form characteristic (White, Swank \& Holt 1982) of
accreting X-ray pulsars, a power law of photon number index $\Gamma $,
E$^{-\Gamma}$, and with high-energy exponential rolloff $E_f$ above a
cutoff energy $E_c$, $e^{-(E-E_c)/E_f}$, was the starting point for
analysis.  To represent cyclotron absorption we employed the
multiplicative profile \[e^{(- \tau \cdot \exp(-(E - E_r )^2/2 \sigma
^2))}, \] where E$_r$ is the centroid, $\sigma $ is the line RMS, and
$\tau $ is the optical depth.  Earlier work in the literature began
with additive and subtractive forms to represent line emission.  With
the general acceptance of the line formation process as resonance
cyclotron scattering of photons from the continuum, multiplicative
profiles were adopted.  With \ginga\ and \bep\ a nearly Lorentzian profile
was also employed:  \[e^{(- \tau \cdot (\sigma \cdot E / E_r )/((E - E_r )^2
+ \sigma ^2)}, \] with $\tau$, $E_r$ and $\sigma$ as above. We report
detailed results only with the Gaussian profile in optical depth, but
note that best-fit values for the line centroid differ only by -5\%
using the Lorentzian profile, and a similar amount for subtractive
profiles.  All results of this analysis are significant also if the
Lorentzian profile is used in place of the Gaussian.  We note that
the employment of any of these profiles is largely historical and heuristic.
Recent calculations (Araya-Gochez \& Harding 2000) indicate a wide
variety of possible line profiles.

Spectra averaged over pulse phase were employed for this work. A major
goal was comparison with historical observations, many of which are
available only in this form.  Although spectra can be strongly phase
dependent in some pulsars, with Her X-1 only modest changes with phase
have been observed for selected parameters, in particular the power law
index change of 30\% (Pravdo et al. 1977) and a 20\% change of the
cyclotron line centroid (Gruber et al. 1980, Voges 1984).  Pulse
phase-resolved spectroscopy will be reported in a later publication.
The observations were collected at various orbital phases (Table 1),
but we expect no effect from this, because a lack of spectral changes
with orbit was reported by Pravdo (1976).  Moreover, the eight \heao\
observations of Soong et al. (1990) showed no orbital variation.  We
performed spectral analysis with the cut-off power law continuum model 
already mentioned, and we tried and rejected another model in current
use, the Fermi-Dirac (Tanaka 1986). 
These forms are used empirically only: we are unaware
of any theoretical forms for the spectral continuum which permit
meaningful comparison with the observed spectra.

\subsection{\xte\ Spectral Analysis}

Data were selected from the five observations such that the source was
uneclipsed either by the primary or the earth and such that characteristic
absorption events (``dips'', $N_H > 10^{21}$cm$^{-2}$) were absent.  A
substantial dead-time correction was applied to the HEXTE data. The
HEXTE background was directly measured by switching the HEXTE look
direction on- and off-source every 16 seconds, while the PCA internal
background was estimated with the \xte\ software tool PCABACKEST and
the strong-source background model.   This model is known to have an
accuracy of the order of a few percent.  Following established
procedure (e.g. Wilms et al. 1999) we tuned the background estimate to
the observed high-energy portion of the data by adjusting a scaling
factor to produce zero PCA net counts above 60 keV.  The RMS correction
for the five observations was 4.4\%.  As a further measure to limit
sensitivity to residual background subtraction errors, PCA data below
3.5 keV and above 25 keV were not used, except as a check.  Within the
3.5 -- 25 keV range the average source spectrum was at least four times
background, and usually much greater.  Even with this favorable
signal-to-background residuals to fits initially showed structure at
the 1\% level.  We found very similar structure in power-law fits to
the Crab spectrum with PCA, and we attributed this to imperfectly
modeled background at the level of accuracy of the background model,
about one milliCrab.  We were able to model the spectrum of the extra
background in the Crab observations as an extremely broad, $\sigma$ =
2.5 keV,  Gaussian line centered at about 5 keV. We then applied it
successfully to remove the dominant residuals in the fits to Her X-1,
and the adjusted constant of normalization again yielded a flux of close 
to one milliCrab, as with the Crab itself.  Except for this extra
background component and for details, our procedure for dealing with
these systematic errors closely follows that described by Wilms et al.
(1999, see Appendix B).  As a practical matter, the HEXTE data above 25
keV are statistically much stronger than the PCA, and very little
statistical weight is gained by adding PCA data above 25 keV.  As a
check, however, we have performed joint fits including PCA data to 60
keV.  The higher-energy PCA data are quite consistent with the HEXTE
data.  In particular, in each observation the PCA independently
confirms the presence of the absorption line at about 40 keV.  The
HEXTE data range extended from 20 keV, chosen to be above the
electronic threshold for all observations, to 100 keV, above which
energy data were discarded because the Her X-1 signal was below
detectability.

Standard PCA energy response matrices produced with version 5.0 ftools
were employed.  These were checked against contemporary Crab observations,
and non-statistical errors of 0.25\% RMS were thus measured.  This error 
was applied to each PCA energy channel as a systematic error.  This 0.25\% 
error represents a refinement of the calibration procedures which earlier
led Heindl et al. (1999) to assign a 1\% systematic error.  
The residuals to the Her X-1 fits with this high-energy cutoff model
for the continuum showed a depression at the cut energy.  This 
was identified as an artifact of this spectral model, which has
discontinuous slope at the cut energy.  This artifact was quite
effectively suppressed by modeling a Gaussian absorption at the cut energy
with optical depth 0.15 and sigma 2.5 keV.  We note that Burderi et al.
(2000), in analyzing \bep\ data for Cen X-3, also found the need to
introduce an empirical rounding function at the cutoff energy with this
continuum.  They used a polynomial rather than a Gaussian form.

With this model and the very low 0.25\% systematic error for PCA, fits
to all five observations produced acceptable chi-squares and a flat
and featureless pattern of residuals.  In particular,
we found no evidence for the 18 keV spectral break reported from
a \bep observation (Dal Fiume et al.  1998). 

Fits to a power law with the Fermi-Dirac (Tanaka 1986) cutoff function, 
\[(1+e^{(E-E_c)/E_f})^{-1},\]
also with fold and cutoff energies but with continuous derivative, were
tried in the hope of modeling the spectral turnover more simply.  Instead,
a pattern of correlated residuals over the entire PCA energy range
resulted, strongly indicating a poor fit and functional mismatch with
the data.  This test with an alternative continuum did show, however, that
best-fit values for the line centroid were even more insensitive to the 
choice of continuum than their 5\% sensitivity to choice of line profile.  

In Figure 1 we show the best fit and residuals for observation 2, which
had the longest duration.  The middle panel shows the large residuals,
particularly at 40 keV, of a fit without cyclotron line, while the
lowest panel shows the acceptable fit when the line is included.  
The displayed PCA data include the 0.25\% systematic error per datum.

In Table 2 we list best-fit parameters and errors from the five
observations for the successful continuum model with power law and 
high-energy cutoff, plus iron K emission and cyclotron scattering
feature.  In the last column of Table 2 are shown results of a
simultaneous fit to all five observations with free normalizations for
each but with the same spectral shape parameters, including line depth.  
For the fit to the combined data the large $\chi^2_r$ of $\sim \! 4.6$ 
is dominated by small but real spectral changes from observation to
observation, primarily in the spectral index and cutoff energy.  We 
estimate the spectrum-to-spectrum variability to be 1.8\% RMS.

Best-fit parameter values in Table 2 were determined by minimizing the
$\chi^{2}$ statistic, using the systematic plus statistical errors for
PCA spectral data and statistical errors only for HEXTE.  The parameter
error estimates were determined from a number of Monte-Carlo
realizations using xspec: starting with a best-fit model we determined
the average and RMS scatter for each parameter.  For the parameters
which define the continuum shape these experimentally-determined errors
averaged a factor 1.8 greater than the values reported directly by
xspec,  while corresponding estimates for the parameters of the
scattering line were larger only by a factor 1.1.  These estimates are
smaller than those obtained by the prescription of Lampton, Margon \&
Bowyer (LMB, 1976), which gave factors averaging 2.2 larger than the
direct xspec estimates.  

Given the satisfactory results obtained with the best-fit spectral
model, we could also search for variability of the individual
parameters of this model.  For the five measured values of each
parameter in Table 2 we calculated averages and chi-squares.  With
either our Monte-Carlo estimates or the LMB errors large chi-squares
for the parameters for the continuum shape strongly indicated very
significant but nevertheless modest variability. Conversely, for the
scattering line parameters both methods indicated stability, with
reduced chi-squares near unity or below.  Reduced chi-squares using the
LMB estimates for the three line parameter errors were all about 0.25,
indicating (P=0.001) that the LMB errors are overestimated.

The parameters which depend most strongly on the correction for PCA
systematic error are the power law index and the iron line parameters.
The cutoff, fold and cyclotron line parameters are determined largely
by the HEXTE data, for which pure statistical errors were employed.

Of special interest is the possible variability of the cyclotron line
centroid.  The five \xte\ centroid values are consistent with no
variability.  The chi-square of 5.7 (4 dof) for the centroid values
permits an extra random variability as high as 1.5 keV (90\%
confidence) in the centroid energy.  

The five \xte\ observations were each made in the first few days of
a Her X-1 35-day main-on state.  Strictly speaking, then, the measured
1.8\% spectral stability applies only to the early part of the main-on.
The spectrum varies somewhat with pulse phase (Soong et al. 1990)
and the pulsation shape is known to vary regularly with 35-day phase
(Scott et al. 1997). So there is reason to expect spectral
variability with 35-day phase.  However, the \heao\ observations (Soong 
et al. 1990), which sampled the main-on state fairly well, showed no 
change of pulse phase-average spectral shape with 35-day phase.

The contemporary observation with \bep\ (Dal Fiume et al. 1998), which
has similar statistical weight to the \xte\ observations, yielded
similar results.  The \bep\ values for the cyclotron line parameters
agree with the \xte\ values to within the errors of Table 2.  There is
reasonable agreement for the iron line parameters.  The continuum
parameters are harder to compare because of the 18 keV break in the
\bep\ power law index, which we do not confirm.  In the next section we
compare these results with observations made well before the launch of
\xte.  This comparison indicates significant changes for the cyclotron
line energy and the cutoff energy of the continuum.  Significant change
was not found in other parameters of the spectrum.

The 41 keV absorption line was observed at high significance in all five
\xte\ observations.  Tr\"{u}mper et al.  (1978), interpreting the
cyclotron feature as emission at 55 keV, reported a possible harmonic
overtone, although this has never been confirmed.  Multiple harmonic
lines have been observed in the spectrum of a few other x-ray pulsars,
most notably 4U0115+63 (Heindl et al. 1999; Santangelo et al. 1999),
which has four or five.  We have searched for a harmonic overtone of
the 41 keV scattering feature by means of a simultaneous fit to the
HEXTE data of all five observations.  The resulting best-fit optical
depth for the overtone at 82 keV is zero, and the 95\% confidence upper
limit is 0.23.  This limit shows that the cyclotron scattering is quite
different in Her X-1 than in 4U0115+63, whose overtones have optical
depths greater than unity.

\subsection{Reanalysis of \heao\ }

To permit more reliable comparison with earlier work we have re-analyzed 
archival \heao\ A2 and A4 spectra using the same high-energy cutoff and 
line profile functions employed here with \xte\ data.  Another important
reason for re-analysis is that considerable work (see Appendix) has gone 
into the calibration of the \heao\ A4 since the results of Soong et al. 
(1990).  For two of the \heao\ observations we were able to fit 
simultaneously 1--18 keV data from an Argon counter of the A2 experiment 
(Rothschild et al. 1978), thereby obtaining a good measure of the power 
law index.  One such fit is shown in Figure 2.  Two sets of residuals
are shown in Figure 2:  the middle panel contains residuals to a fit
without the cyclotron line and clearly shows a pronounced deficit at 35 
keV;  the bottom panel contains acceptable residuals to a fit which
includes the line.

The \heao\ spectral parameter values and errors shown in Table 3 are
from this re-analysis.  By comparison with Soong et al. (1990), values
differ somewhat.  The spectral index and cutoff energy are smaller by
20\%, the fold energy is 10\% larger, the line centroid (3\% smaller)
is nearly the same, and the line width is smaller by 30\%.  The change
of line width is probably not significant, given the resolution and
statistics.  As with the \xte\ data, we tested the sensitivity of the
cyclotron line centroid value to different line profiles, as well as
the two continuum models.  We found changes only of the order of 1 keV,
even less than with \xte, Errors were estimated as described here for 
\xte.  They are a factor of 2-3 larger than those reported by Soong et al.
(1990), who used a simpler procedure to estimate errors.

The most interesting result from the reanalysis are the \heao\ values
for the cutoff, which lie between 16 and 17 keV, considerably smaller
than the 24 keV reported as a best-fit for \heao\ on-pulse minus
off-pulse spectra by Gruber et al. (1980), and also less than the 21
keV assumed by Soong et al. (1990).  These two authors did not have the
use of the \heao\ A2 data, which powerfully constrains the value of the
spectral index.  Indeed, the values of the spectral index, cutoff and
fold are very highly correlated in the A4 data, because these data have a
lower threshold at 13 keV.  The correlation has the sense that
overestimating the cutoff energy requires the fit to compensate by
moving the index to a larger value and the fold energy to a smaller
value.  When we fit to the A4 data alone, the index moved to a value
above unity, the  cutoff energy moved to 18 keV, and the fold energy to
13 keV.  A cutoff energy at 21 keV, as used by Soong et al. (1990), 
requires a chi-square higher than this A4-only best-fit by $\sim \! 4$.  
Such a difference is reasonably consistent with small changes in
the response matrix, background subtraction, and functional forms from
those used earlier.  We thus believe that the earlier cutoff values of
Gruber et al. (1980) and Soong et al. (1990) were considerably
overestimated, largely due to the limited spectral range of the A4
data.  We note that the \heao\ A2/A4 data sets are very well matched to
the \xte\ PCA/HEXTE data sets in spectral range and statistical
significance.  Therefore comparisons of spectral parameters between
observations performed with these missions is much more reliable than
between other pairs of missions.

\subsection{Long Term Changes}

The values for the line centroid in the present work are roughly
consistent with measurements of values near 44 keV obtained between
1993 and 1995 with {\it CGRO}/BATSE (Freeman et. al 1996).  The \xte,
\bep\ and BATSE values differ from all earlier measurements, which
centered near 35 keV (Kunz 1995).  Earlier data with good statistics
and reliable energy calibrations include those from the \heao\ A4
experiment (Soong 1988), the GRIS germanium detector (Tueller et al.
1984), the HEXE experiment aboard the Russian platform {\it MIR\/}
(Kunz 1995), and \ginga\ (Mihara 1995).  However, functional forms for
the representation of the continuum and line profile have been various,
making comparison somewhat uncertain.  In Table 3 we have collected
reported results for continuum (power law with a high energy exponential 
cut off) and cyclotron line parameters.  The iron line parameters,
which did not change sensibly, are ignored.

The cyclotron energy between 1976 and 1996 is charted in Figure 3.
Clearly, the data before 1991 are consistent with an average value near
35 keV and the data after 1991 are consistent with a higher average of 41 keV.
When we analyze the parameters in Table 3 for the formal significance
of a change of average value (a step function in time) using the
F-test, we obtain the average values and significances shown in Table 4. 
The large reduced chi-squares are perhaps inevitable given  calibration
errors and somewhat differing functional forms for fits, as well as
the possibility of variability on any time scale.  Formal statistical
results based on such high chi-squares are not valid. However the
F-statistic, which is based here on the comparison of chi-squares for
the null hypothesis of no change, $\chi ^2 _n$, and for a change between
1991 and 1994, $\chi ^2 _f$, should be more robust and provide at least
an upper limit to the the null hypothesis (no change) probability.
With this statistic  the line centroid has in fact a very low significance
$\leq 1.8 \times$ 10$^{-11}$ for no change.  Probably significant is a
change of the cutoff energy, with P$_{null} \leq $ 0.01.  Possibly
significant at 90\% confidence are small changes of the power law index
and fold energy, and a doubling of the scattering line width.  If the
Lorentzian line profile is substituted in the analysis, the changes are
slightly less significant.

The observations of continuum cutoff energy relied almost entirely on
the \heao\ and \xte\ measurements.  The cutoff energy could not be
measured with the many balloon experiments shown in Table 3 and Figure
3 because of atmospheric absorption below 25 keV.  The MIR/HEXE
experiment had an electronic threshold at 25 keV and was therefore
insensitive to cutoff energy.  The \bep\ fit was made with a more
complicated continuum model with two inflection points near 20 keV, so
could not be used.  Finally, errors were not stated for the
\ginga\ observation. 
The most striking feature of the behavior of the cutoff energy is its
apparent close proportionality to the centroid energy, shown in Figure
4.   The dashed line displays proportionality with the best-fit factor
of 2.03.  The correlation coefficient for these data is 0.97, with a
null (no correlation) probability of only $1.1 \cdot 10^{-6}$.

We conclude that a historical change has occurred in the
values of at least two characteristics of the Her X-1 spectrum: the
cyclotron scattering energy and the spectral cutoff energy of the
power-law continuum.  

\section{Discussion}

Models for the overall X-ray spectra and pulse shape from X-ray pulsars 
(Kirk et al. 1986; M\'{e}sz\'{a}ros \& Nagel 1985) have had only moderate
success in predicting the observed continuum spectral shapes.  We
concentrate therefore on the interpretation of the cyclotron line,
especially in its bearing on the pulsar magnetic field.  We discuss
also the correlation of the continuum cutoff with centroid.  The
absence of sizeable change in the other spectral parameters should
perhaps not be surprising given the likelihood that an important
controlling factor in the spectral shape is the rate of energy release,
and that this varied only over a factor of two in the joint \xte\ and
\heao\ data sets, for which we can speak most authoritatively.  The
stability of the parameters of the fluorescent iron emission at 6.5 keV
and their similarity to the values obtained with OSO-8 (Becker et al.
1977) and \bep\ (Dal Fiume et al.  1998) indicate that the geometry of
the reprocessing site, usually thought to be at the magnetopause (e.g.
McCray et al. 1982), has not changed dramatically.

\subsection{Cyclotron Line Centroid Variability with the \xte\ Dataset}

The absorption feature is widely regarded to result from resonance
scattering of electrons in a magnetic field (Tr\"{u}mper et al. 1978)
of intensity $\sim \! 3 \times 10^{12}$ Gauss (Voges et al. 1982), given
the earlier measurements of 35 keV for the centroid.  These five
\xte\ observations give a new average value of $3.6 \times 10^{12}$
Gauss and can be used to set limits to the range of magnetic field
strengths sampled.  These \xte\ data are consistent with no
variability of line centroid to within 3\% RMS.  The field strength at the
scattering site is therefore also stable to a precision of 3\%.  The
magnetic moment can be determined from the inferred field strength,
to be $1.8 \times 10^{30}$ Gauss-cm$^3$, assuming  magnetic dipole
structure, scattering at the polar surface of the neutron star and a 
10 km radius.  For scattering somewhere above the surface, this limit
constrains the range for the average height of scattering to within 180
meters (90\% confidence) in a dipole field, and requires a somewhat
higher magnetic moment.

Next, we test the \xte\ data against a more specific process.  There
seems to be no general agreement on the structure of the radiating
region of the polar cap of an X-ray pulsar.  If a shock forms in the
flow above the pole BAK (1991) predict an accretion mound of sizeable
extent.  Other theorists, however, most recently Bulik et al. (1995), place 
the site of deceleration of the accretion flow at the base of the neutron
star atmosphere, with a scale height of less than a meter.  The present
results are consistent with the latter scenario, but they do not
constrain the accretion mound scenario unless this is hundreds of
meters or more in height.  However, we note that if the observed line
width of 5 keV or 12\% is due largely to a mix of magnetic field
strengths in a dipole field, then the scattering occurs over a range of
the order of 1 km in height.

BAK estimate an accretion mound  height of 650 meters for Her X-1, with
a direct proportionality of this height to mass transfer rate
(luminosity).  Although BAK state that the radiation escapes from the
mound at its base, Mihara et al. (1999) have had some success modeling
centroid variability in several pulsars with intensity under the
assumption that the scattering occurs at the top of the mound. Mihara
et al. (1999) do not find such a clear pattern, however, for Her X-1.
As shown in Figure 5, the present \xte\ data also rule out the
BAK/Mihara centroid-intensity variability, to better than 3$\sigma$
confidence.  The \xte\ observations span a 2--30 keV flux range of
(4.5--8)$\times$10$^{-9}$ erg-cm$^{-2}$-s$^{-1}$, for which
BAK/Mihara predict a change of 7 keV. In addition, there is  no
evidence in any of the \xte\ observations for photoelectric absorption, 
therefore an actual correlation could not have been disguised
by unrelated obscuration of the X-ray emitting region.

The fairly tight \xte\ constraints on variability of line centroid,
interpreted as a limit on the range for height of scattering in a
dipole magnetic field, give some preference for emission at the stellar
surface.  We note, however, that on the pulsation time scale the change
of line centroid with pulse phase reported by Soong et al. (1990) could
be produced in the BAK/Mihara picture if the average height of
scattering changes with viewing angle, i.e., pulse phase.  It has not
yet been decided observationally, however, whether it is the centroid
that changes more with pulse phase or the line profile more generally.

\subsection{Variability on long Time Scales}

This new observation of a long-term change of cutoff energy closely 
proportional to the change of scattering energy suggests a common factor 
in producing both spectral features.  This controlling factor, undoubted
in the case of the scattering feature, is almost certainly the magnetic
field strength.  
In a study of twelve X-ray pulsars with cyclotron lines Makishima et al.
(1999) demonstrated a convincing correlation between the cutoff energy
of the continuum and the line centroid.  Their functional relation is
a power law with index 1.4, thus cutoff $\sim$ centroid$^{1.4}$.
Could this population characteristic also apply to changes in time
with a single pulsar, as in this case?  We find (Figure 4) that a power
law relation with index 1.4 fits the \xte\ and \heao\ data much worse
than simple proportionality.  However, we note that the data in Figure
4a of Makishima et al. (1999) seem reasonably consistent also with
direct proportionality, so that there may indeed be some correspondence.
 
These long-term spectral changes could result from a change of the
height of scattering, although we can note that in this case the
BAK/Mihara scenario would require \heao\ fluxes two to three times what
was observed, and thus can be ruled out.

Brown \& Bildsten (1998) have reconsidered the structure of the
settling mound addressed earlier by BAK and have developed a model with
a more compact structure, of the order of 100 meters in height.  In
calculating the hydrostatic pressure of the accreted settling matter in
this structure, they obtain a pressure comparable to the magnetic
pressure, which distorts and spreads the magnetic field lines, thus
lowering the magnetic field intensity.  They give no time scale for
significant change of this structure.  The observed limit of 5 years to
the period of change may be applicable to their model.

Ruderman et al. (1998) have a model of neutron stars in which motions
of superfluid neutron vortices may alter the magnetic field in the
core, which can result in movements of the stellar crust which in turn
alter the surface magnetic field.  The observed change of line centroid
could thus in principle be related to a change of magnetic field
in response to a change of structure. A structural change, in turn,
would very likely produce a timing event.  However, it is not known
whether such a crustal change could produce a timing glitch large
enough to be seen in the presence of accretion timing noise.

\section{Conclusions}

We have shown convincing evidence for a 23\% historical change in the
cyclotron line energy in the spectrum of Her X-1 through comparison of
\xte\ observations with earlier observations.  For this purpose archival 
\heao\ spectra were re-analyzed, using modern tools and spectral forms for
consistency.  On the other hand, the five \xte\ observations, which
span about two years, show no change, with a limit of about three
percent.  We also demonstrate a historical change of continuum cutoff
energy which is closely proportional to the cyclotron centroid
energy.  This is the first good evidence to connect the cutoff energy,
which is common to accreting pulsars, to the magnetic field.  These
\xte\ data do not support a particular accretion mound scenario.

\section{Acknowledgements}

This work was supported in part by NASA under contracts NAS8--27974, 
NAS5-30720 and grant NAGW--449. The authors at T\"{u}bingen gratefully
acknowledge support from German grants DLR 50 00 9605 and
DAAD/Staubert.  Valuable comments from the referee led to a much
improved treatment and presentation.

\newpage

\begin{center}
{\bf Appendix: Calibration of the \heao\ A4 LED detectors}
\end{center}

The two Low Energy Detectors (LED's) of the High-Energy X-ray and
Low-Energy Gamma-Ray Experiment (A4) aboard \heao\ were thin NaI
scintillators of area 103 cm$^2$ each, energy range 12--180 keV, and
energy resolution 25\% FWHM at 60 keV.  Many spectral results were
published from these detectors, but only a very brief description of
the energy calibration and monitoring procedures has been reported
(Jung 1986).  For energy calibration the LED detectors were provided
with $^{241}$Am calibration sources whose events were recognized
through a special pulse shape.  But this calibration required special
operating modes that were employed infrequently.  Moreover, unexpected
changes in the crystal optical interfaces following launch left the
absolute energy calibration with these sources uncertain.  Therefore an
alternative procedure was devised which relied on the correct
identification as well as monitoring of spectral lines in the
instrument internal background.  This background resulted largely from
radioactive daughter isotopes produced by spallation interactions of
cosmic rays and geomagnetically-trapped protons with the iodine nuclei
in the detector.  Most daughters are unstable and beta decay, producing
a continuum in energy within the detector.  But an alternative channel
for positron decay is capture by the nucleus of an atomic electron in
the K shell.  This K-capture produces characteristic radiation whose
total energy results from the fluorescent radiation of the electron
cloud as it reconfigures as well as the nuclear deexcitation gamma ray,
if any.  A number of such lines were identified in the LED detectors
(Gruber et al.  1988).  The most useful line for calibration and
monitoring was from $^{125}$I (half life 60 days), for which there was
a gamma with energy 35 keV and a total of 32 keV from the electronic
cascade.  Because all the charge was collected by the electronics in
less than the several microsecond pulse shaping time, the summed pulse
corresponding to 67 keV energy loss was observed.  The identification
of this isotope and line in the A4 data was quite secure, because the
60-day buildup of the activity at the beginning of the mission was
clearly observed.

The most important complication for energy calibration for the A4
detectors was the energy-dependent efficiency of light production by
the inorganic phosphor (Adams \& Dams 1970).  Light production in the
crystal lattice is stimulated by fast electrons resulting from the photon
interaction. This process apparently has a maximum efficiency at
electron energy of 10--20 keV (e.g. Zerby, Meyer \& Murray 1961).  This
variation in light output with energy for a HEXTE detector is shown in
Figure 4 of Wayne et al. (1998).  The energy-dependent efficiency was
carefully calibrated prior to launch of the \heao\ using radioactive
sources.  This calibration curve was then used to calculate the light
production for a $^{125}$I decay, whose energy is shared among several
electrons ranging from a few keV to tens of keV.  The light production
was calculated to be equivalent to light originating from a single
photon of 62.7 keV, not 67 keV.  Use of the former number successfully
resolved a problem in the interpretation of LED spectra: a strong spurious 
40 keV emission line which appeared in the spectra of the Crab, Cyg
X-1, and the diffuse X-ray background!  Reassignment of the calibration
energy from 67 keV to 62.7 keV completely removed this artifact in all
these spectra .  From the sensitivity to this feature alone we estimate
the absolute accuracy of the A4 LED calibration as 2\%.  Other aspects
of the calibration were discussed most fully in several UCSD PhD
dissertations:  effective area, Nolan (1982); angular response, Knight
(1981); gain variability, monitoring and correction, Jung (1986).

\newpage

\begin{center}
{\bf REFERENCES}
\end{center}

\begin{description}
\item[] Adams, F. and Dams, R., 1970, "Applied Gamma-Ray Spectrometry", 
(Oxford:Pergamon), pp 55-57.
\item[] Araya-Gochez, R. A., \& Harding, A. K., 2000, Ap. J., 544, 1067.
\item[] Becker, R. H. et al., 1977, Ap. J., 214, 879.
\item[] Brown, E. F. and Bildsten, L., 1998, Ap. J., 496, 915.
\item[] Bulik, T. et al., 1995, Ap. J., 444, 405.
\item[] Burderi, L. et al., 2000, Ap. J., 530, 429.
\item[] Burnard, D. Arons, J., and Klein, R., 1991, Ap. J., 367, 575
\item[] Dal Fiume, D. et al., 1998, A\&A, 329, L41.
\item[] Freeman, P. E. et al., 1996, in ``Proceedings of the 3rd Huntsville
Symposium'', eds Kouveliotou, Briggs \& Fishman, (New York:AIP), p. 172.
\item[] Gruber, D. E.  et al., 1980, ApJ, 240, L127.
\item[] Gruber, D. E., et al., 1989, in ``High-Energy Radiation Background
in Space'' eds. Rester \& Trombka, (New York:AIP), p 232.
\item[] Gruber, D. E., et al., 1997, in ``Proceedings of the 4th Compton
Symposium'', eds Dermer, Strickland and Kurfess (New York:AIP), p. 744.
\item[] Gruber, D. E., et al., 1998, in ``The Active X-Ray Sky'' eds Scarsi,
Bradt and Giommi (Amsterdam:Elsevier), p 174. 
\item[] Gruber, D. E., et al., 1999, in ``Proceedings of the 2nd Integral
Workshop'', in press.
\item[] Heindl, W., et al., 1999, ApJ, 521, L49.
\item[] Isenberg, M., Lamb, D. Q., and Wang, J. C. L., 1998, ApJ, 505, 688.
\item[] Jahoda, K., et al., 1996, Proc. SPIE, 2828, 59.
\item[] Jung, G. V., 1986, unpublished dissertation, UCSD.
\item[] Kirk, J. G., 1986, A\&A 169, 259.
\item[] Knight, F. K., 1981, unpublished dissertation, UCSD.
\item[] Kunz, M., unpublished dissertation, 1995, Univ. T\"{u}bingen.
\item[] Lampton, M., Margon, B., \& Bowyer, S., 1976, Ap. J., 208, 177.
\item[] Makishima, K., et al., 1999, Ap. J., 525, 978.
\item[] McCray, R., et al., 1982, Ap. J., 262, 301.
\item[] M\'{e}sz\'{a}ros, P., and Nagel, W., 1985, Ap. J., 299, 138.
\item[] Mihara, T., 1995, unpublished dissertation, University of Tokyo.
\item[] Mihara, T., et al., 1998, Adv. Space Res., 22, 7, 987.
\item[] Nolan, P. L., 1982, unpublished dissertation, UCSD.
\item[] Orlandini, M. et al., 1999, Proceedings of 1999 Bologna Meeting,
in press.
\item[] Pravdo, S. H., 1976, unpublished dissertation, University of Maryland.
\item[] Pravdo, S. H., et al., 1977, ApJ, 216,L23.
\item[] Pravdo, S. H., et al., 1979, M. N. R. A. S., 188, 5.
\item[] Rothschild, R. E., et al., 1978, Space Sci. Instr., 4, 265.
\item[] Rothschild, R. E., et al., 1998, ApJ, 496, 538.
\item[] Ruderman, M., et al., 1998, ApJ, 492, 267.
\item[] Santangelo, A., et al., 1999, ApJ, 523, L85.
\item[] Scott, D. M., et al., 1997, in ``Proceedings of the Fourth 
Compton Symposium, eds Dermer, Strickland and Kurfess (New York, AIP), 748. 
\item[] Soong, Y., 1988, unpublished dissertation, 1998, University of
California, San Diego.
\item[] Soong, Y., et al., 1990, Ap. J., 348, 641.
\item[] Tanaka, Y., 1986, in IAU Colloq. 89, eds D. Mihalas and K. H. Winkler,
(New York, Springer), 198.
\item[] Tr\"{u}mper, J., et al., 1978, Ap. J. (Letters), 219, L105.
\item[] Tueller, J., et al., 1984, Ap. J., 279, 177.
\item[] Voges, W., 1984, unpublished dissertation, LMU M\"{u}nchen.
\item[] Voges. W., et al., 1982, Ap. J., 263, 803.
\item[] Wayne. L., et al., 1998, Nucl, Inst \& Meth. in Physics Res., Sect A, 
411, 351.
\item[] White, N. E., Swank, J. H., and Holt, S. S., 1983, Ap. J., 270, 711.
\item[] Wilms. J., et al., 1999, Ap. J., 522, 460.
\item[] Zerby, C. D., Meyer, A. \& Murray, R. B., 1961, Nucl. Inst. Meth. 12, 115.
\end{description}

\newpage

\setcounter{figure}{0}

\figcaption{ (Top panel) Fit with high-energy cutoff continuum and
Gaussian absorptive line profile to PCA and HEXTE data from 1996 July
in units of detector counting rate versus energy.  Except above 70 keV,
where adjacent channels have been grouped for display, original detector
energy channels are shown.  These oversample the actual detector energy
resolution by factors of three to eight.  (Middle panel) Residuals to
fit {\em without} absorption line, showing strong and correlated
deviations, particularly at 40 keV, the line energy. (Bottom panel)
Residuals, now acceptable, to fit which includes the absorption line.
PCA errors include a 0.25\% systematic error, which dominates the
statistical errors. HEXTE markings, x for Cluster A and o for cluster
B, have been omitted in the top panel for clarity.  Error bars are
one sigma.
}

\figcaption{Spectral fit and residuals to \heao\ data of 1978 Feb 24.
The three panels are as in Figure 1.  The 1-16 keV data are from an
Argon-filled detector of the A2 experiment, and the 12-100 keV data
marked with circles and crosses are from the two Low Energy Detectors
of the A4 experiment.  Errors (one sigma) are purely statistical.
Original energy channels are shown, and these oversample the detector
resolution by factors of two to four.  Other than a negative feature at
4.5 keV, the fit is acceptable.  The best-fit cyclotron energy is 33.2
keV and the cutoff energy 15.9 keV.  
}

\figcaption{
History of centroid energy of absorption line.  Errors are one sigma,
as estimated by the original investigators or in this paper for \xte\
and \heao.  A pronounced change is evident following the \ginga\
measurement in 1990.  Dashed lines indicate the averages for the
earlier and later data.  A similar and proportional change of the
cutoff energy is also quite significant.  The internal consistency of
the \xte\ data indicates that the errors have not been underestimated.
Observations are listed by date and experiment in Table 3.  
}

\figcaption{
The value of the centroid energy of the cyclotron scattering versus the 
cutoff energy of the spectral continuum, with 1-$\sigma$ errors.
Correlation is formally highly significant, and the dashed line shows
the best-fit proportionality with constant 2.03.  Errors are determined
from the 68\% joint confidence regions.  A power law dependence with
index 1.4 (dot-dashed line), as was observed in a population study by
Makishima et al. (1999), fits quite a bit worse.  
}

\figcaption{
Measured line centroids versus 20-60 keV flux obtained with HEXTE-only
datasets, with 1-$\sigma$ errors.  The line shows expected dependence
predicted by Mihara (1999) from  the model of BAK (1991), in which the
mean luminosity corresponds to a height of 1 km above the polar cap for
the scattering.  The predicted dependence is not confirmed.  
}

\newpage

\begin{table}
\caption{Outline of Observations}
\begin{center}
\begin{tabular}{clcc} \hline \hline
Observation & Date & Duration in sec & Binary Phase Range$^{\dagger}$\\ \hline
1 & 1996 Feb &  5000 & .17-.23\\
2 & 1996 July & 28000 & .63-.93, .12-.46 \\
3 & 1996 Oct & 5800 & .06-.08, .49-.52 \\
4 & 1997 Sept & 20000 & .13-.38 \\
5 & 1997 Nov & 18000 & .21-.40, .68-.74 \\ \hline
\multicolumn{4}{l}{$^{\dagger}$ Zero phase is mid-eclipse, or superior conjunction}\\
\end{tabular}
\end{center}
\end{table}

\newpage

\begin{table}
\caption{Results of Spectral Fitting, PCA $<$25 keV plus HEXTE Data}
\begin{center}
\begin{tabular}{|l|ccccc|c|} \hline \hline

date  &  1996 Feb   & 1996 July & 1996 Oct   & 1997 Sept & 1997 Nov  & all\\[-0.04in] \hline \hline
\multicolumn{6}{|l|}{2-30 keV flux, units 10$^{-9}$ keV cm$^{-2}$ sec$^{-1}$} & \\ \hline
flux & 7.84 & 7.27 & 5.33 & 6.83 & 4.44 & - \\ \hline \hline
\multicolumn{6}{|l|}{power law with high-energy cut-off continuum} & \\[-0.04in] \hline \hline
index$^{\dagger}$ &  0.855(4)$^{\ddagger}$ & .862(4)   &  .841(5)  &  .901(4) & 0.852(4) & .841(2) \\[-0.04in]
cutoff     &   19.83(11) & 20.23(8)  &  19.10(12) & 20.27(9) & 19.50(10) & 19.60(9) \\[-0.04in]
fold       &   10.87(22) & 10.65(15) &  10.02(19) & 10.49(19) & 10.08(16) & 10.79(9) \\[-0.04in] \hline

Fe line & & & & & &  \\[-0.04in]
eqw      &   .154(19)   & .190(18)    &  .219(24)   & .140(16)   &  .188(23) & .171(9) \\[-0.04in]
centroid   &   6.625(31) & 6.518(24) &  6.524(25) & 6.490(31) & 6.393(33) & 6.535(15) \\[-0.04in]
sigma      &   .31(4)  & .37(3)  &  .34(3)  & .30(4)  & .48(4) & .36(2) \\[-0.04in] \hline

scat. line & & & & & & \\[-0.04in]
depth      & 0.80(5) & 0.74(3) & 0.63(3) & 0.66(5) & .67(4) & .75(2) \\[-0.04in]
centroid   &   40.9(5) & 40.8(3) &  39.4(7) & 40.1(4) & 40.0(6) & 41.0(2) \\[-0.04in]
sigma      &   4.5(5)  & 4.9(3)  &  3.8(7)  & 4.7(4)  & 3.7(6) & 4.9(2) \\[-0.04in] \hline
$\chi^2_r$ &  1.10     & 0.99      & 0.94  & 0.98 & 1.08 & 4.56 \\[-0.04in]
\hline \hline
\multicolumn{7}{l}{$^{\dagger}$ parameters have units keV except for index and depth, both dimensionless.} \\[-0.04in]
\multicolumn{7}{l}{$^{\ddagger}$ 68\% statistical errors in parentheses apply to the last significant figure(s)} \\[-0.04in]
\end{tabular}
\end{center}
\end{table}

\newpage

\begin{table}
\begin{center}
\caption{Historical Observations of Parameters for Cut-off Continuum}
\begin{tabular}{llcccccc} \hline \hline
year   & Data Set   & $\Gamma $ & E$_c$     & E$_f$     & $\tau $           & E$_r$     & $\sigma $ \\ \hline
+1900  &            &           &       keV & keV       &                   & keV       &   keV     \\ \hline
77.671 & BalloonHEXE &           &           & 12.2(9)   & 3.2(6)$^\dagger $ & 36.8(14)  & 1.3       \\
78.151 & \heao\     &           & 16.4(9) & 10.8(4) & 1.20(55)           & 34.0(8) & 2.9(13)  \\
78.161 & \heao\     &           & 16.6(10) & 11.1(5) & 1.48(98)           & 33.9(8) & 2.5(14)  \\
78.164 & \heao\     &           & 16.1(15) & 11.7(7) & 1.78(81)          & 34.5(9) & 3.4(13)  \\
78.622 & \heao\     &  .943(14) & 16.0(4) & 10.9(4) & 1.16(37)          & 33.2(9) & 2.6(13)  \\
78.630 & \heao\     &  .912(12) & 16.8(4) & 11.4(4) & 0.87(9)          & 34.6(9) & 4.5(11)  \\
78.638 & \heao\     &           & 16.9(5) & 9.8(4) & 0.68(17)          & 33.2(9) &   \\
80.726 & GRIS       &           &           & 9.9(16) &                   & 35.4(21)  & 3.59(188) \\
87.528 & HEXE       &           &           &  10.6(10) & 3.0(18)           & 33.8(18)  &           \\
87.618 & HEXE       &           &           &   9.9(10) & 1.6(10)           & 32.3(23)  &           \\
88.565 & HEXE       &           &           &  10.8(20) & 4.5(34)           & 35.0(21)  &           \\
90.070 & \ginga\    &    0.93   & 19.2      & 15.3      &                   & 36.8      & 6.75      \\
94.370 & BATSE    &      &    &   &       & 44(1)      & 2.3(10)      \\
96.085 & \xte\      &   .855(4) & 19.83(11) & 10.87(22) & 0.80(5)           & 40.9(5) & 4.5(5)  \\
96.562 & \bep\      &           &           &           & 0.73(3)           & 40.3(2)   & 6.3(4)    \\
96.567 & \xte\      &   .862(4) & 20.23(8) & 10.65(15) & 0.74(3)           & 40.8(5) & 4.9(3)  \\
96.762 & \xte\      &   .841(5) & 19.10(12) & 10.02(19) & 0.63(3)          & 39.4(13) & 3.8(7)  \\
97.688 & \xte\      &   .901(4) & 20.27(9) & 10.49(19) & 0.66(5)           & 40.1(7) & 4.7(4)  \\
97.860 & \xte\      &   .852(4) & 19.50(10) & 10.08(16) & 0.67(4)           & 40.0(6) & 3.7(6)  \\ \hline
\multicolumn{8}{l}{$^\dagger $ Equivalent width (keV) for subtractive Gaussian}
\end{tabular}
\end{center}
\end{table}

\newpage

\begin{table}
\caption{Average Values of Spectral Parameters in Table 3}
\begin{center}
\begin{tabular}{lcccccc} \hline \hline
 & $\Gamma $ & E$_c$ & E$_f$ & $\tau $ & E$_r$ & $\sigma $ \\ \hline
 & &       keV & keV       &          & keV   &   keV \\ \hline \hline
Pre-1991 & .927(8) & 16.52(17) & 10.84(20) & 0.90(11) & 33.67(25) & 3.32(34) \\
Post-1991 & .864(10) & 19.89(21) & 10.40(16) & 0.71(3) & 41.02(40) & 4.83(36) \\
change   & -.063(13) & 3.37(27) &  -0.44(25) & -.19(11) & 7.36(47) & 1.51(49) \\ 
ratio & .932(13) & 1.204(18) & .959(23) & .789(102) & 1.218(15) & 1.455(184) \\ \hline
$\chi ^2$ (no change), dof & 210.7/7 & 281.5/10 & 35.2/15 & 47.7/16 & 831.6/20 & 35.2/12 \\ 
$\chi ^2$ (change), dof & 128.7/5 & 97.1/8   &  29.6/13  & 41.5/14 & 54.7/18 & 28.1/10 \\
P$_{null}$ & 0.29 & 0.014 & 0.33 & 0.38  & $2.3 \cdot 10^{-11} $ & .33 \\ \hline \hline
\end{tabular}
\end{center}
\end{table}

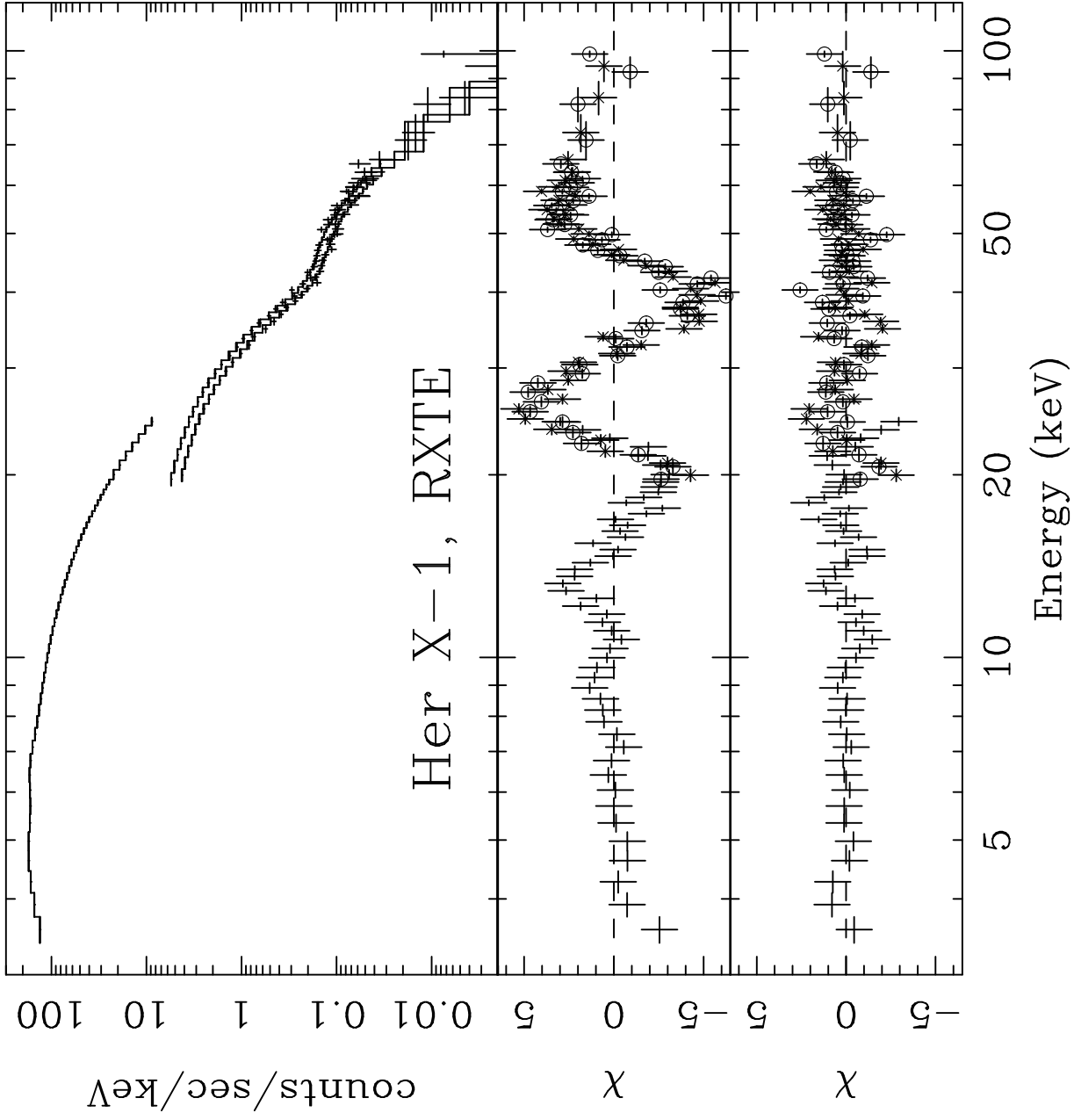
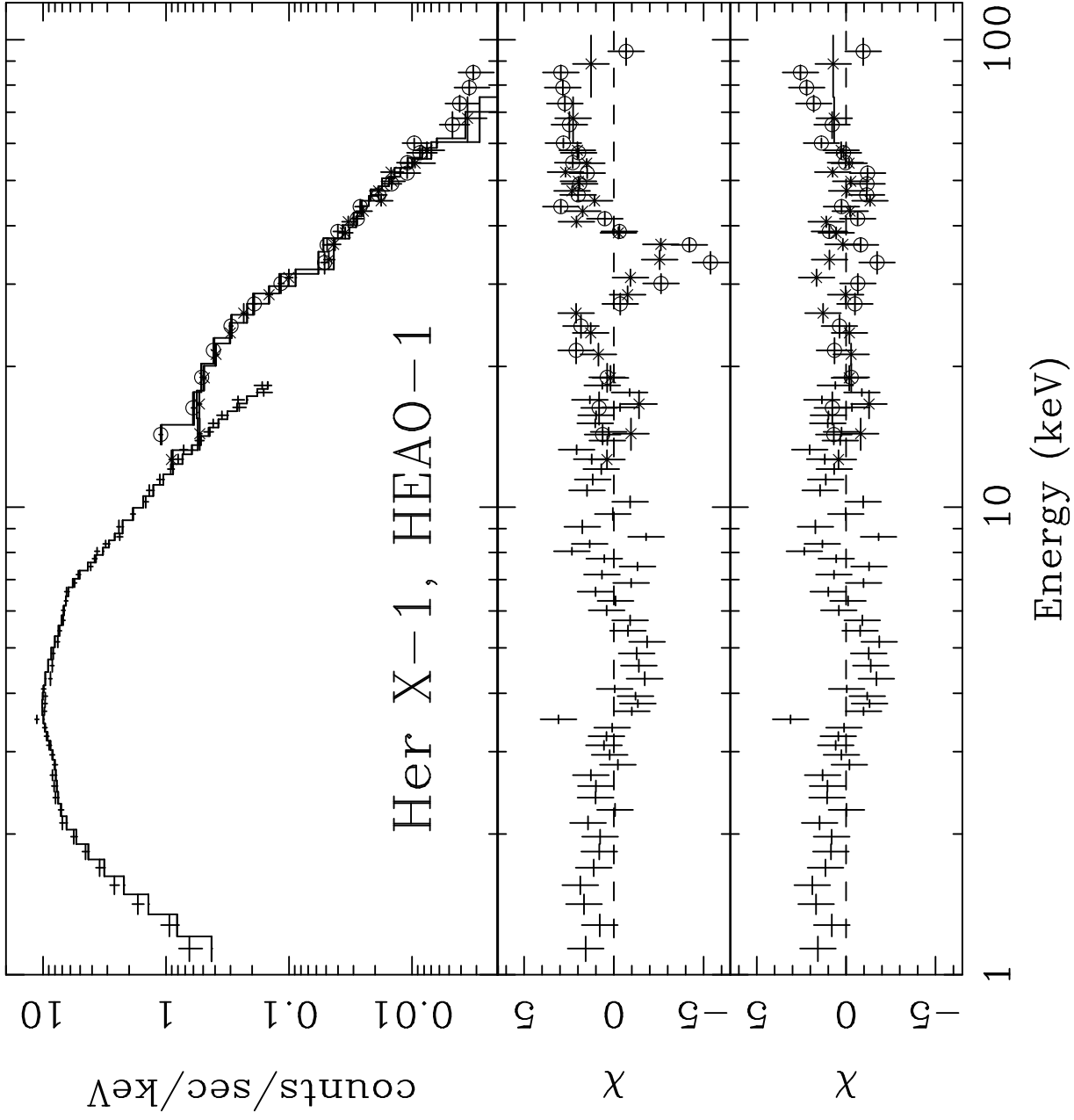
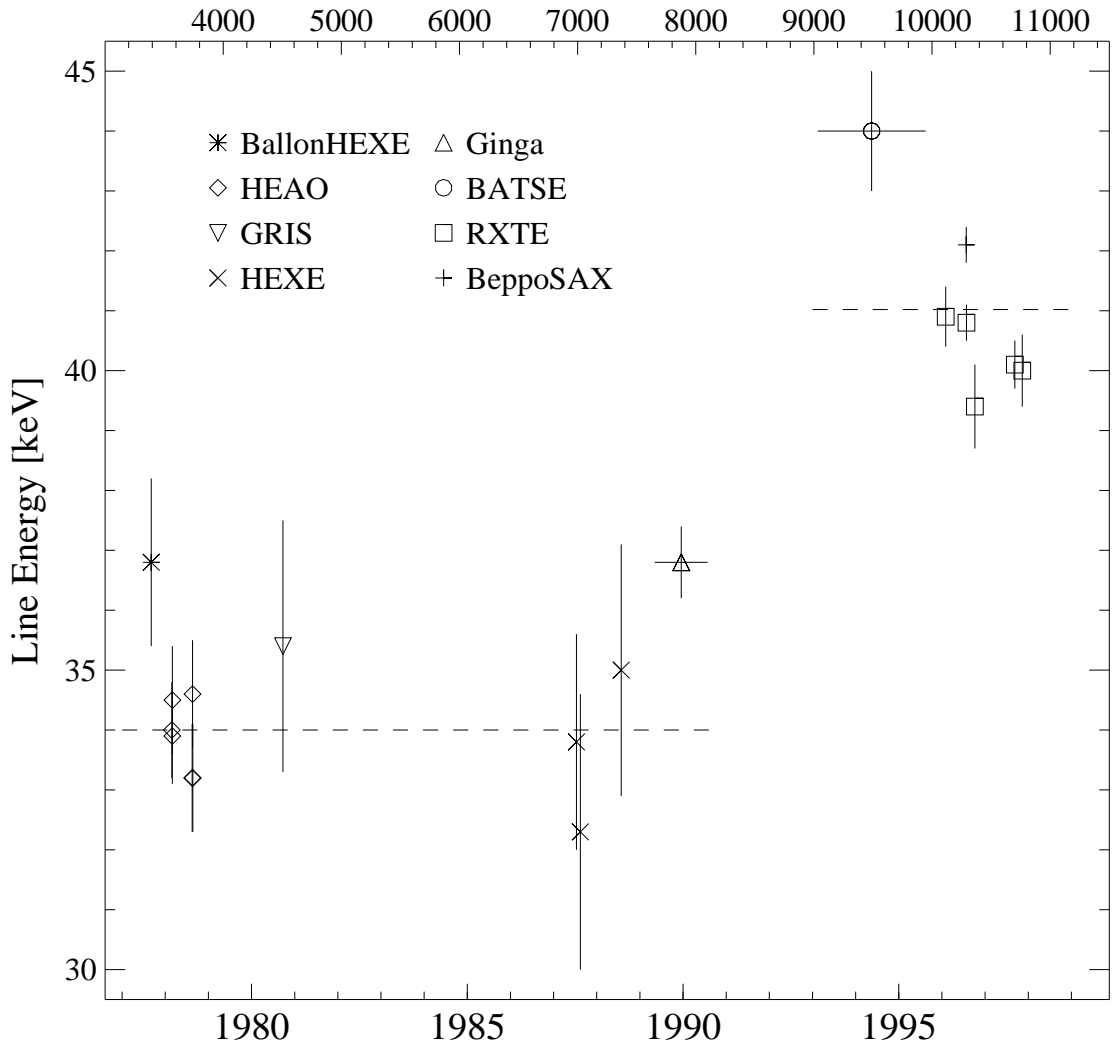
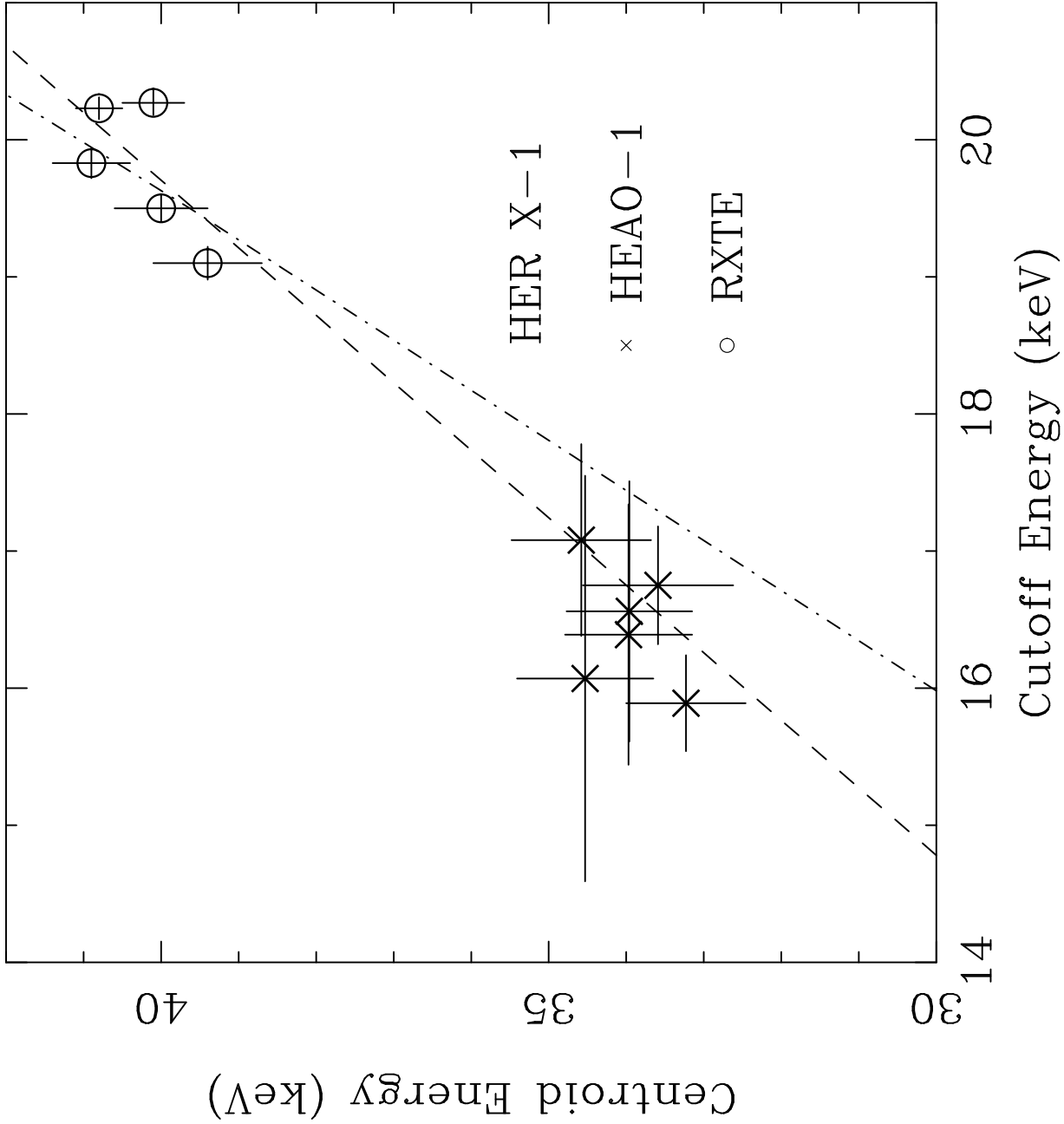
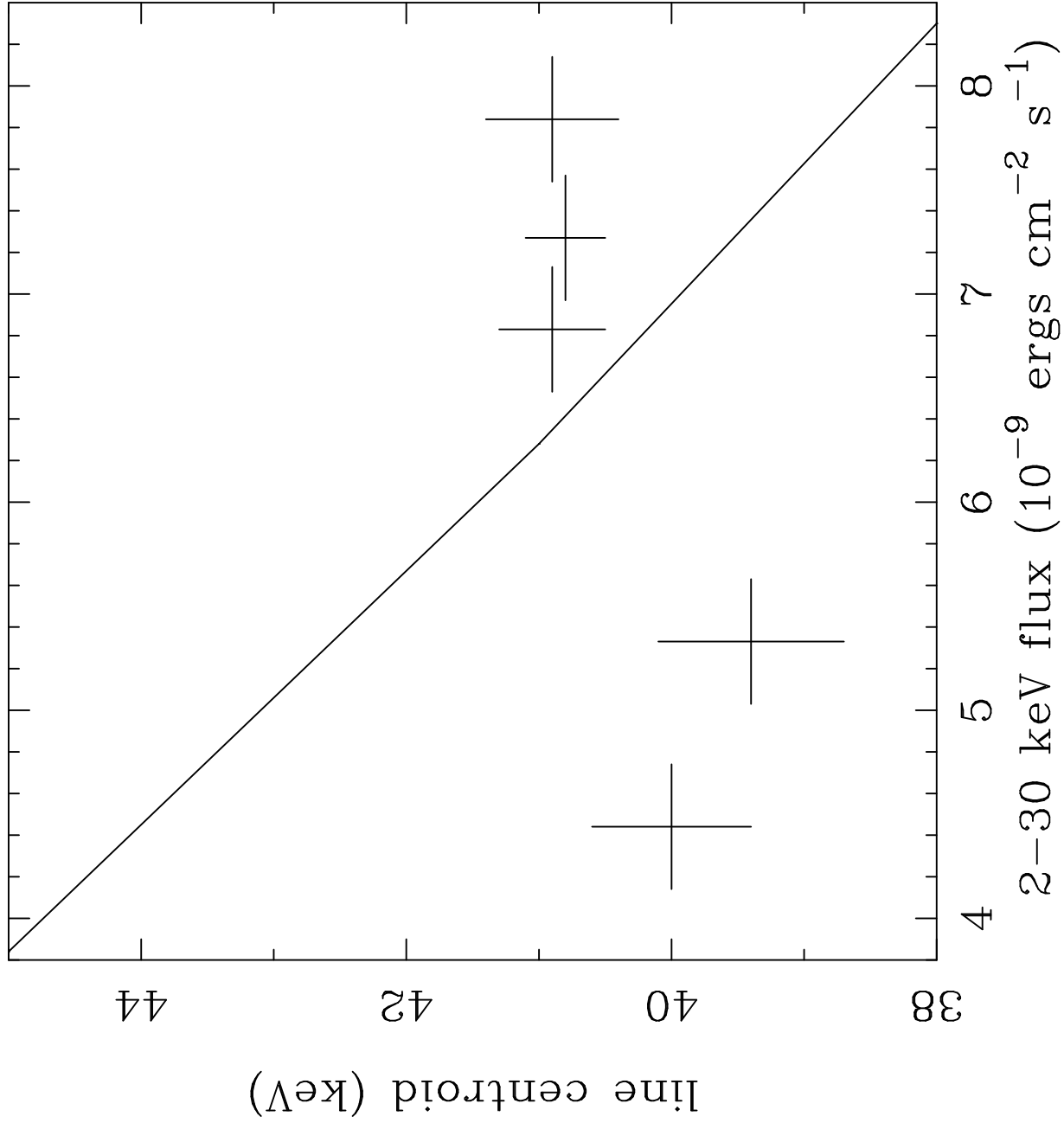
\end{document}